\documentclass[journal,comsoc]{IEEEtran}

\usepackage{blindtext, graphicx}
\usepackage[backend=bibtex,style=ieee,sorting=ynt]{biblatex}
\usepackage[T1]{fontenc}
\usepackage{amsmath}
\usepackage[cmintegrals]{newtxmath}

\usepackage{tikz}   
\usepackage{pgfplots}
\usepackage{siunitx} 
\usepackage{pgfplots}
    \usetikzlibrary{
        pgfplots.colorbrewer,
    }
    \pgfplotsset{
        cycle list/.define={my marks}{
            every mark/.append style={solid,fill=\pgfkeysvalueof{/pgfplots/mark list fill}},mark=*\\
            every mark/.append style={solid,fill=\pgfkeysvalueof{/pgfplots/mark list fill}},mark=square*\\
            every mark/.append style={solid,fill=\pgfkeysvalueof{/pgfplots/mark list fill}},mark=triangle*\\
            every mark/.append style={solid,fill=\pgfkeysvalueof{/pgfplots/mark list fill}},mark=diamond*\\
        },
    }

\begin{document}

\title{Deep Learning-Based MIMO Communications}

\author{Timothy~J.~O'Shea,~\IEEEmembership{Senior Member,~IEEE,}
        Tugba~Erpek,~\IEEEmembership{Member,~IEEE,}\\
        and~T.~Charles~Clancy,~\IEEEmembership{Senior Member,~IEEE}
\thanks{Authors are with the Bradley Department of Electrical and Computer Engineering at Virginia Tech and DeepSig, Arlington, VA, 22203 USA e-mail: (oshea, terpek, tcc)@vt.edu.}}

\maketitle

\begin{abstract}
We introduce a novel physical layer scheme for single user Multiple-Input Multiple-Output (MIMO) communications based on unsupervised deep learning using an autoencoder.  This method extends prior work on the joint optimization of physical layer representation and encoding and decoding processes as a single end-to-end task by expanding transmitter and receivers to the multi-antenna case.  We introduce a widely used domain appropriate wireless channel impairment model (Rayleigh fading channel), into the autoencoder optimization problem in order to directly learn a system which optimizes for it.  We considered both spatial diversity and spatial multiplexing techniques in our implementation. Our deep learning-based approach demonstrates significant potential for learning schemes which approach and exceed the performance of the methods which are widely used in existing wireless MIMO systems.  We discuss how the proposed scheme can be easily adapted for open-loop and closed-loop operation in spatial diversity and multiplexing modes and extended use with only compact binary channel state information (CSI) as feedback.
\end{abstract}


\section{Introduction}

\IEEEPARstart{M}{ultiple}-Input Multiple-Output (MIMO) wireless systems are widely used today in 4G cellular and wireless local area network systems to increase throughput and coverage by exploiting the multipath characteristics of the channel.  By encoding information across multiple antenna elements using spatial diversity or multiplexing range or throughput can be improved in various channel conditions.  Traditional MIMO communication schemes are divided into two categories which are either open-loop (without any Channel State Information (CSI) information at the transmitter), or closed-loop systems (with CSI at the transmitter fed back from the receiver). These MIMO schemes rely on rigid analytically obtained encoding/beamforming and decoding schemes for these tasks and are in general not known to be optimal. Moreover, closed-loop spatial multiplexing techniques rely on CSI and channel estimation and sending that information to the transmitter results in estimation and feedback (i.e. quantization) errors which further complicates the ability of these schemes to perform optimally. 

In recent years, applications of Machine Learning (ML) are proposed to be integrated in communication systems. \cite{heath} proposes to use the Support Vector Machines (SVM) to enable machine learning classification to determine Modulation and Coding scheme. \cite{jeon} presents a blind detection framework that performs data symbol detection without explicitly knowing CSI at a receiver. These approaches use ML as an additional functional block to perform a certain task in the conventional communication systems.    

Recent results in physical layer learning for the Single Input Single Output (SISO) channel \cite{intromlcomsys} have shown that autoencoders \cite{goodfellow2016deep} can readily match the performance of near-optimal existing baseline modulation and coding schemes by learning the system during training.  In this paper, we extend the use of autoencoders in SISO systems to MIMO systems. Our main contributions in this paper are as follows:

\begin{itemize}
    \item We introduce a scheme which can combine many MIMO tasks into a single end-to-end estimation, feedback, encoding, and decoding process which can be jointly optimized to maximize throughput and minimize bit error rate for specific channel conditions.  We believe this joint system optimization process has the potential to provide significant gains in comparison with the current day system which is often optimized in a more modular fashion. Moreover, the performance of the developed model can continuously be improved by training for longer periods since generating input data (i.e., symbols) is straightforward.  
    \item We demonstrate that it is possible to achieve and exceed the performance of the conventional spatial diversity MIMO systems using Deep Learning (DL) techniques and autoencoders. First, we simulated the performance of a Space Time Block Code (STBC) scheme for 2x1 system which was proposed in \cite{alamouti}. Next, we developed an end-to-end learning MIMO communication system as a Deep Neural Network (NN) that is trained to learn the transmitted symbols that go through a Rayleigh fading channel. We compare the Bit Error Rate (BER) with that of an autoencoder system. The results show that the autoencoder system exceeds the performance of the STBC code when the Signal-to-Noise Ratio (SNR) value is above approximately 15 dB.
    \item We demonstrate that it is possible to achieve and exceed the performance of the conventional linear-precoding based closed-loop single user MIMO spatial multiplexing system using DL and autoencoders. We simulated the performance of a closed-loop MIMO system that has perfect CSI at the transmitter. We used Singular Value Decomposition (SVD)-based precoding at the transmitter to send the symbols. The receiver decodes the received signal and estimates the transmitted symbols. Next, we developed a Deep NN model that uses the CSI at the transmitter during training. The autoencoder learns the CSI along with the transmitted symbols and uses this information to eliminate the interference at the receiver and estimate the transmitted symbols in high accuracy. The results show that the autoencoder system exceeds the performance of the SVD-based MIMO spatial multiplexing system for all SNRs.
    \item We modify our NN implementation to consider the practical challenge in real world systems of the bandwidth constrained feedback channel for providing CSI. We consider the case where a compact $v$-bit representation of the CSI is available at the transmitter instead of perfect real-valued CSI. In other words, we introduce quantization error to the system for each channel entry and learn $2^v$ discrete channel states for which the information can be encoded. The results show that quantization of the CSI actually improves the performance of our system for certain values of $v$, providing even lower bit error rates across the whole range of SNRs.
\end{itemize}
The rest of this article is organized as follows. Section \ref{background} provides background information for spatial diversity and multiplexing techniques we considered in this paper as well as the autoencoder systems. Section \ref{approach} explains the autoencoder model and the DL algorithms we used in detail. The simulation results are included in Section \ref{results}. Section \ref{discussion} contains a discussion of open problems and next steps. Section \ref{conclusion} provides the conclusions.    

\section{Background} \label{background}

\subsection{Spatial Diversity}
MIMO spatial diversity techniques; i.e., STBC codes, are used to increase robustness and extend coverage of a system. The STBC scheme for a 2x1 MIMO system introduced in \cite{alamouti}, known as Alamouti code, achieves the same performance as Maximal Ratio Receiver Combining (MRRC) when the transmit power is doubled. With this approach, at a given symbol period, two different symbols are transmitted from the two antennas. If we denote the symbol transmitted from the first antenna as $s_1$ and the second antenna as $s_2$, during the next symbol period, $-s_{2}^*$ is transmitted from the first antenna and $s_{1}^*$ is transmitted from the second antenna where $^*$ denotes complex conjugate form. Note that there is no improvement in throughput with this approach since it still takes two time slots to transmit two symbols.   

At the receiver, the signals at antenna ports 1 and 2 can be written as $r_1 = h_1s_1 + h_2s_2 + n_1$ and $r_2 = -h_1s_{2}^* + h_2s_{1}^* + n_2$, respectively where $h_1$ and $h_2$ are the channel variables with circularly symmetric complex Gaussian entries with zero mean and unit variance and $n_1$ and $n_2$ are noise terms with also zero mean and unit variance Gaussian variables. A maximum likelihood or Minimum Mean Square Error (MMSE) detector is used to estimate $s_1$ and $s_2$.     

The STBC for 2x1 MIMO systems were generalized to $N$ transmit and $M$ receive antennas in \cite{STBCGen}. However, we will focus on the performance of a 2x1 system for performance evaluations in this paper. 

\subsection{Spatial Multiplexing}

MIMO spatial multiplexing techniques are used to increase the single and multi-user throughput \cite{telatar}, \cite{mac}. We consider a single user MIMO system with $N_t$ antennas at the transmitter and $N_r$ antennas at the receiver. The transmitter sends $(N_t \times 1)$ symbols out and the signal received at the receiver can be modeled as $\textbf{y} = \textbf{H} \textbf{s} + \textbf{n}$ where $\textbf{H}$ is an $(N_r \times N_t)$ channel matrix with circularly symmetric complex Gaussian entries of zero mean and unit variance, $\textbf{s}$ is an $(N_t \times 1)$ vector with transmit symbols with a total power constraint of $P$ such that $\mathbf{E}[\textbf{s*} \textbf{s}]\leq P$. $\textbf{n}$ is an $(N_r \times 1)$ vector which is the additive white Gaussian noise noise at the receiver with $\mathbf{E}[\textbf{n}\textbf{n*}]=\sigma^2 \textbf{I}_{N_r \times N_r}$. We assume the noise variance, $\sigma^2$, is $1$ in this paper.

Each receive antenna receives signals from all the transmit antennas. Different precoding/beamforming approaches have been developed to eliminate the interference at each antenna port and demodulate the signals correctly in the literature \cite{MIMOSurvey}. In this paper, we consider a closed-loop MIMO system which uses SVD-based precoding technique \cite{tse}. The channel matrix, $\textbf{H}$, can be written as $\textbf{H}= \textbf{U} \Lambda \textbf{V*}$  where $\textbf{U}$ and $\textbf{V}$ are $(N_r \times N_r)$ and $(N_t \times N_t)$ unitary matrices, respectively. $\Lambda$ is a diagonal matrix with the singular values of $\textbf{H}$. In order to eliminate the interference at each antenna, the channel can be diagonalized by precoding the symbols at the receiver and decoding at the receiver using the CSI. In this model, the received signal can be written as $\tilde{\textbf{y}} = \Lambda \tilde{\textbf{x}} + \tilde{\textbf{n}}$ where $\tilde{\textbf{x}} = \textbf{V} \textbf{x}$, $\tilde{\textbf{y}} = \textbf{U*} \textbf{y}$ and $\tilde{\textbf{n}} = \textbf{U*} \textbf{n}$. The distribution of $\tilde{\textbf{n}}$ is the same as $\textbf{n}$ with $\tilde{\textbf{n}} \sim \mathcal{N}(\mu, \, \sigma^{2} \textbf{I}_{N_{r}})$. A Zero Forcing (ZF) or MMSE equalizer can be used to estimate the transmitted symbols. 
The closed-loop MIMO systems consume the system resources with sending pilot signals from the transmitter to the receiver for CSI estimation and sending the CSI back. We also considered the system performance when the quantization error is introduced to the system to decrease the amount of feedback by representing each entry of CSI in a $v$-bit compact form while sending back to the transmitter.     

\subsection{Channel Autoencoders}

\begin{figure}[h]
\centering
\includegraphics[width=0.5\textwidth]{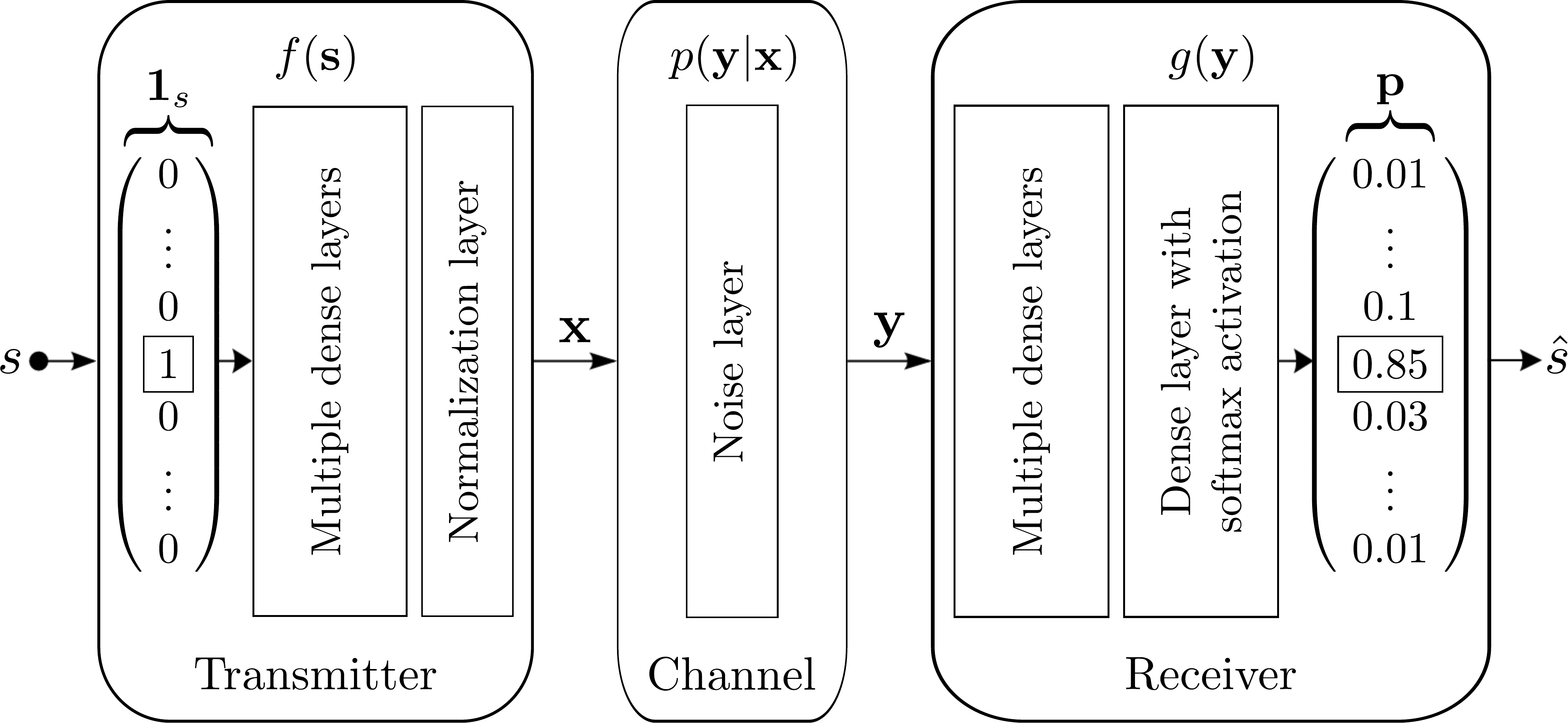}
\caption{SISO channel autoencoder system.}
\label{fig:cae}
\end{figure}

Recent results in physical layer learning for the Single Input Single Output (SISO) channel \cite{intromlcomsys} have shown that autoencoders \cite{goodfellow2016deep} can readily match the performance of near-optimal existing baseline modulation and coding schemes by using an autoencoder to jointly learn the system. More recently this approach has also been shown to work very effectively over the air \cite{dlota}, as well.

The idea of the channel autoencoder \cite{learningtocommunicate, intromlcomsys} applies deep unsupervised learning with a reconstruction-based loss function to jointly optimize encoding, decoding, and signal representation over some impaired communications channel.  An illustration of the basic channel autoencoder system is shown in Figure \ref{fig:cae}.  

This approach is appealing as it finds solutions for modulation and forward error correction which rival today's best designs for small code word sizes over existing channel impairment models, and offers a method to learn solutions over channel impairments for which no optimal solution is known or a compact analytic representation may be difficult to express or optimize for when considering real world effects.  Another benefit of such systems is that the computational complexity of the learned encoder and decoder modules can in many cases be of lower computational complexity than existing methods, leading to potential power savings when deploying such systems efficiently.  

\section{Technical Approach} \label{approach}

In the SISO channel autoencoder, $k$ bits are encoded into $n$ sequential time samples to be transmitted over some communications channel such as a Rayleigh fading wireless channel.  When noisy versions of these samples are received by a decoder, an estimate for $\hat{k}$ is produced to best reconstruct the original information.

In this work, we consider the MIMO case where $k$ bits ($s$) are encoded to form $m_t$ parallel transmit streams of $n$ time-samples. These streams then undergo some form of multi-antenna mixing channel and arrive at a receiver to produce $m_r$ receive streams, each also of length $n$ time samples and decoded to produce an estimate for $\hat{s}$.  Thus there are $m_r \times m_t$ unique pairwise channels, each with some random impairment due to wireless propagation. 

This is the model conventionally used in wireless systems, where a MIMO transmission may optimize to varying numbers of transmit and receiver antennas, and symbols such as in an Alamouti code, may be repeatedly encoded over adjacent samples in time using different spatial modes.  

By varying the number of information bits $k$, the transmit streams $m_t$, the receive streams $m_r$ and the number of time samples $n$ the information is encoded over, this model may be readily adapted to different information rates.  By selecting how many bits are encoded into a given space-time block, and over what time and antenna dimension the space-time block spans encodings which optimize for high rate dense spatial multiplexing as well as lower rate diversity improvement can be obtained readily using the same optimization problem.  

\begin{figure}[h]
\centering
\includegraphics[width=0.5\textwidth]{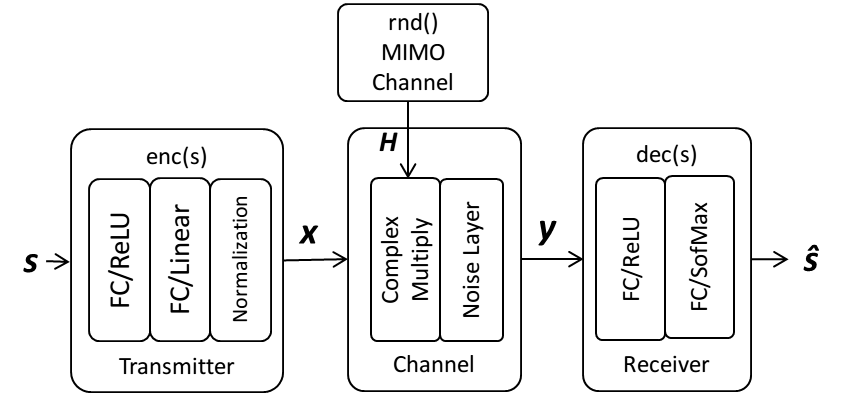}
\caption{MIMO Channel Autoencoder (No CSI).}
\label{fig:mimocae1}
\end{figure}

To realize and optimize such a MIMO channel autoencoder we present the open-loop MIMO system shown in Figure \ref{fig:mimocae1} where we leverage multi-layer NNs to provide the mappings from information bits or codewords to digital samples to transmit, and from received digital samples back to information bits or codewords.  Without any knowledge of CSI, this is a simple way to learn an encoding, we will use this approach directly below to learn a solution to the 2x1 channel which can be compared directly to Alamouti in Section \ref{nocsi}.  The encoder and decoder both use batch normalization between each layer, and $s$ is introduced using an embedding layer.  

\begin{figure}[h]
\centering
\includegraphics[width=0.5\textwidth]{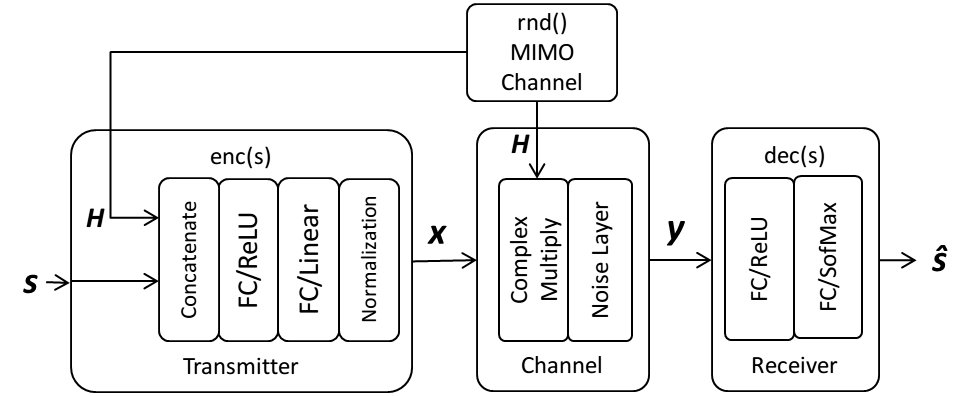}
\caption{MIMO Channel Autoencoder (Perfect CSI).}
\label{fig:mimocae2}
\end{figure}

To consider the case where the transmitter has perfect CSI of the current fading conditions between each antenna, we consider an extension to this scheme, shown in Figure \ref{fig:mimocae2}, which concatenates $s$ to $H$ prior to encoding for the channel.  This simple step is sufficient to learn greatly improved performance competitive with current day baseline schemes such as ZF, as we discuss in Section \ref{csi}.

\begin{figure}[h]
\centering
\includegraphics[width=0.5\textwidth]{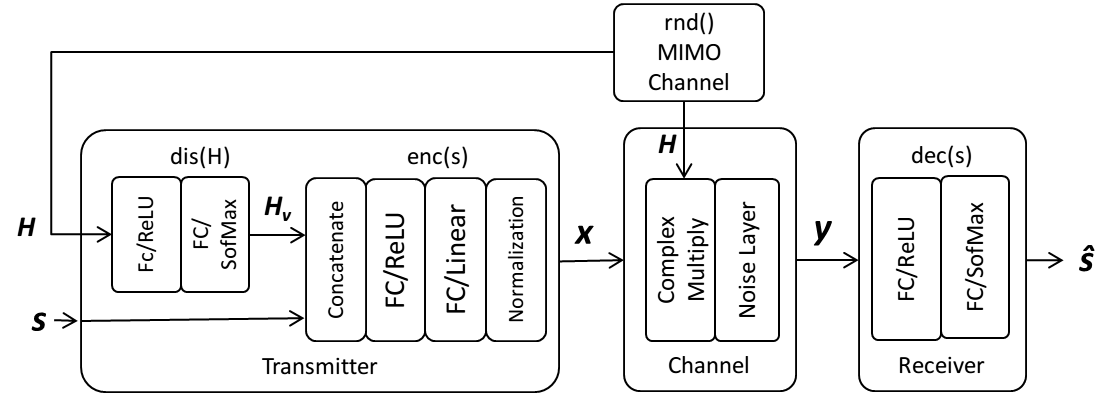}
\caption{MIMO Channel Autoencoder (v-Bit Discrete CSI).}
\label{fig:mimocae3}
\end{figure}

Additionally, we consider the case of CSI with quantized values, encoding knowledge of the real-valued $H$, into a $v$-bit binary vector $H_v$.  In real world systems this is important as the ability to most compactly transmit CSI to mobile devices is paramount.  In Figure \ref{fig:mimocae3} we illustrate an extension to consider discrete-valued channel states, where we introduce a $v$-bit, $2^v$ valued classification problem on $H$ to form $H_v$ prior to concatenation and encoding with $s$.  

\begin{figure}[h]
\centering
\includegraphics[width=0.5\textwidth]{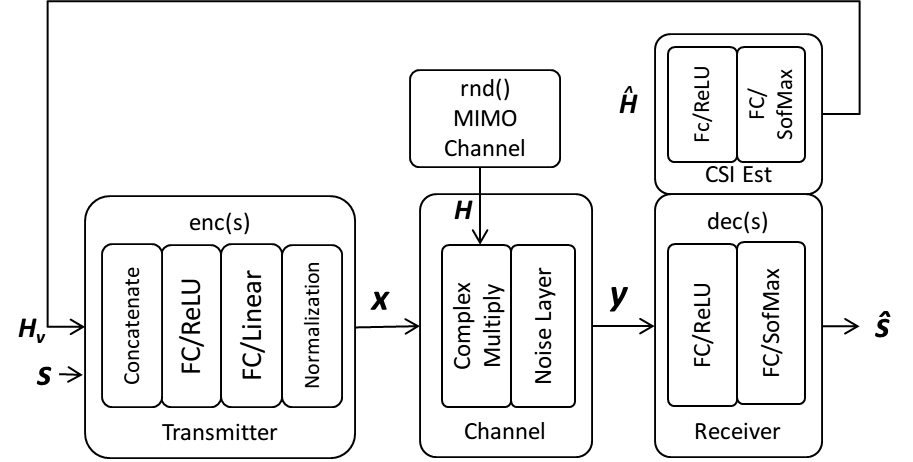}
\caption{Deployment Scheme for v-Bit CSI MIMO.}
\label{fig:mimocae4}
\end{figure}

Finally, we consider the deployment scenario for the $v$-bit CSI scheme derived above.  In this case, a system may estimate the channel $\hat{H}$ at the receiver, producing $\hat{H_v}$ based on estimation, and transmit that information back to the transmitted for its use in encoding.   This process is simplified for the symetric channel where $H_v$ can be assumed equal in both directions and transmission of the estimate may not be needed.  The scheme is illustrated in Figure \ref{fig:mimocae4} showing a full closed-loop system which can jointly learn a method for compact binary CSI feedback, encoding, and decoding of information over the MIMO fading channel.  This system may be trained as shown in Figure \ref{fig:mimocae3}, but the learned network mapping to a compact CSI form of $H_v$ is simply wholesale to the receiver where an estimated channel response $\hat{H}$ may be obtained using either traditional estimation approaches or NN-based regression.

\subsection{Optimization Process}


In our optimization process, we represent $s$ as a $2^k$ valued integer of codeword indices which may be transmitted by the system, each encoding $k$ bits.  In the network, we present this as a one-hot input vector of length $2^k$ with a single non-zero value of 1 for the desired codeword, and the output as a soft-max classification task which approximates the probability of each code word.  
In such classification tasks, a categorical cross-entropy loss function ($\ell_{CE}$) may be readily used for optimization using gradient descent to select network parameters.  
In this case $\ell_{CE}$ is given by 

\begin{equation}
    \ell_{CE}(y,\hat{y}) = \frac{-1}{ \left| y \right| } \sum_{i=0}^{ \left| y  \right| -1} \left( y_i log(p_i) + (1-y_i)log(1-p_i) \right)
\end{equation}
where our output prediction $\hat{y}$ is equivalent with $p_i$.

Using a form of stochastic gradient descent (Adam \cite{adam}), weight updates are computed based on the loss gradient using back-propagation.  In this case, we iteratively compute a forward pass: $\hat{y} = f(y,\theta) $, and a backward pass, $\frac{\partial \ell}{\partial \theta} = \frac{ \partial \ell_{CE} ( y, f(y,\theta) }{ \partial \theta } )$ where network layer weights are given by $\theta$ and a weight update takes the form of $\delta w = -\eta \frac{\partial \ell}{\partial \theta}$ and $\eta$ represents a (possibly time varying) learning rate.

\subsection{Channel Simulation and Network Architecture}

Core to the ability to optimize such an end-to-end system is to accurately model or represent the MIMO channel effects within the network transfer function $\hat{s} = f(s,\theta)$.  To that end, we introduce several new layers which simulate real world impairments and design constraints for each forward and backward pass.  

After the encoding to $m_t$ complex valued transmit streams, a transmit block-code tensor of shape [batch size, $m_t$, 2, $n$] is formed for transmission where the third dimension has the real and imaginery values.  To simulate MIMO propagation we use several custom layers to model the domain enumerated below.

\begin{itemize}
    \item enc:  Learned Encoder: $ s \mapsto x$
    \item rnd:  Random $m_r \times m_t$ channel response $H$
    \item mul:  Complex matrix multiplication of $x$ with $H$
    \item norm: Normalize average power
    \item awg:  Additive gaussian noise ~ $N(0,\sigma)$
    \item dec:  Learned Decoder: $ r \mapsto \hat{s} $
\end{itemize}

In terms of these basic operations, we can express the full network $f$ as follows for the open loop MIMO encoding case:

\begin{equation}
    f(s,\theta) = \text{dec}(\text{awg}(\text{norm}(\text{mul}(\text{enc}(s),\text{rnd}())),\sigma))
\end{equation}
and the for the closed-loop MIMO encoding case as: 

\begin{equation}
    f(s,\theta) = (\lambda H, \text{dec}(\text{awg}(\text{norm}(\text{mul}(\text{enc}(s,H),H)),\sigma)))(\text{rnd}())
\end{equation}

Using this formulation, forwards and backwards gradient passes can readily be computed on $f(s, \theta)$.  In the backwards pass, the awg function becomes the identify function (it is used only for forwards passes).  While the normalization module enforces a constant average power, the noise deviation $\sigma$ may be easily adjusted at training or test time to simulate varying levels of SNR.

\section{Simulation Results} \label{results}

In this section we train the autoencoder based learned encoding model described above and evaluate the Bit Error Rate (BER) performance over a range of SNRs and compare the performance to widely used baselines used under different channel conditions. Matlab is used to simulate the conventional MIMO systems for both spatial diversity and multiplexing. Keras with Tensorflow backend is used for the DL autoencoder implementation using a GPU backend.

We consider comparisons to two different configurations; the 2x1 Alamouti STBC intended to provide spatial diversity based range extension and performance improvements, along with the 2x2 MIMO system that uses spatial multiplexing. QPSK modulation scheme is used to modulate the input bits for the Matlab simulations and Rayleigh fading is used as the channel model for both of the simulations. 

\subsection{Spatial Diversity} \label{nocsi}

\begin{figure}[h]
\centering
\includegraphics[width=0.4\textwidth]{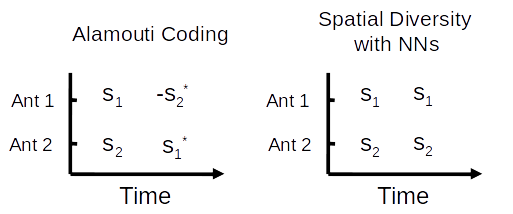}
\caption{Symbols Transmitted in Alamouti Scheme and Neural Networks.}
\label{fig:alamoutiexp}
\end{figure}

Alamouti coding groups symbols to two sequential time-slots. It sends two different symbols from each antenna in the first time slot. The modified versions of the symbols are sent in the second time slot as shown in Figure \ref{fig:alamoutiexp}. An MMSE receiver is typically used to demodulate the bits. 

In the autoencoder implementation, we allow $s$ to describe an equal number of bits (4) which describes 2-bit symbols, an equal information density to the Alamouti code described above and used in our simulation.  Similarly the autoencoder maps these bits to two subsequent complex real valued time-slots equivalent to the Alamouti transmit density.  It is assumed that no CSI is available at the transmitter (i.e. open-loop MIMO).

\begin{figure}
 \centering
 \begin{tikzpicture}
  \begin{axis}[
            cycle list/RdGy-6,
            mark list fill={.!75!white},
            cycle multiindex* list={
                RdGy-6 \nextlist
                my marks \nextlist
                [3 of]linestyles \nextlist
                very thick \nextlist
            },  
	xmode=linear,
	ymode=log,
	xlabel=Signal to Noise Radio (dB), 
	ylabel=Bit Error Rate (BER),
	title={2x1 Spatial Diversity Code Comparison},
	grid=both,
	minor grid style={gray!25},
	major grid style={gray!25},
	width=0.85\linewidth,
    legend pos=south west,
	no marks]
    \addplot table[x=snr,y=ber,col sep=comma]{2x1ae.csv};
    \addplot table[x=snr,y=ber,col sep=comma]{2x1a.csv};
    \legend{{2x1 AE No CSI},{2x1 Alamouti}
    }
  \end{axis}
 \end{tikzpicture}
 \caption{Error Rate Performance of Learned Diversity Scheme.}
 \label{fig:2x1ber}
\end{figure}
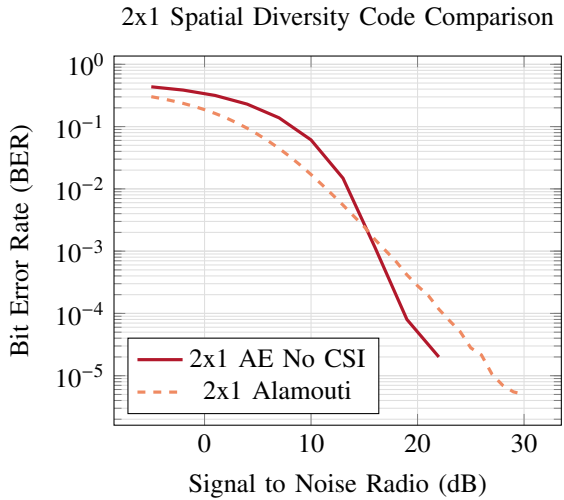

Figure \ref{fig:2x1ber} shows the SNR vs BER results for both Alamouti STBC coding and autoencoder system. In our preliminary results we obtain slightly better performance from the Alamouti STBC at low SNR and improved performance from the autoencoder above around 15dB.  These results are achieved without significant hyper-parameter tuning or long training runs, and we believe these results ultimately hold significant potential for improvement and serve as a lower bound for the performance by which similar networks can achieve.

\begin{figure}[h]
\centering
\includegraphics[width=0.5\textwidth]{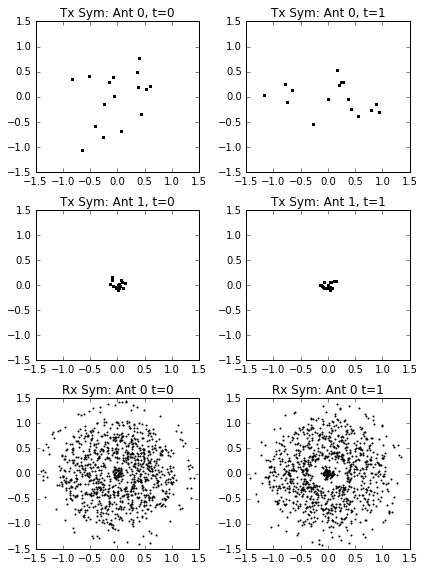}
\caption{Learned 2x1 Scheme, Random Channels.}
\label{fig:2x1consta}
\end{figure}

\begin{figure}[h]
\centering
\includegraphics[width=0.5\textwidth]{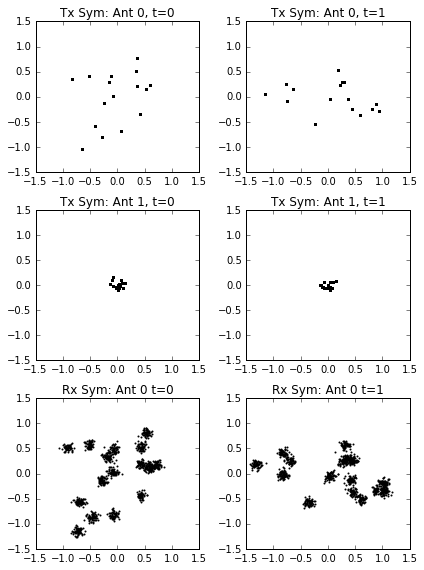}
\caption{Learned 2x1 Scheme, Diagonal Channel.}
\label{fig:2x1constb}
\end{figure}

Inspecting the transmitted and noisy received constellation points at high SNR (20dB) for the learned 2x1 system provides some insight into the achieved learning.  In Figure \ref{fig:2x1consta} we show the constellation points over numerous random draws from the channel matrix $\textbf{H}$.  It appears here that the scheme has learned to achieve its average transmit power by an uneven power distribution between the two antennas in this case, yielding a sort of super-position coding scheme.  In Figure \ref{fig:2x1constb} we show the same constellations but for a constant channel matrix $\textbf{H}$ where the superposition of transmission schemes from antennas 1 and 2 can be seen on the receive side.  This is somewhat similar to the superposition-style constellations learned in \cite{intromlcomsys} for the SISO interference channel.

\subsection{Spatial Multiplexing} \label{csi}
We investigated the performance for two different configurations; when there is perfect CSI and when there is quantized CSI at the transmitter for spatial multiplexing.

\subsubsection{Perfect Channel Information at the Transmitter}

We simulate a 2x2 closed-loop MIMO system with a single time-slot using both a conventional communication system and an autoencoder approach. It is assumed that the channel estimation is perfect and there is no feedback error while sending the CSI back to the transmitter. Rayleigh fading is used as the channel mode and equal power is used at each antenna during transmission for both the baseline and autoencoder methods.

The transmitter uses CSI for SVD-based precoding to diagonalize the channel and eliminate interference at the receiver for the baseline method. There are 100 subframes in the simulation each of them with 1000 symbols. It is assumed that the coherence time of the channel is equal to the duration of a subframe. Thus, the same channel estimation is used for 1000 symbols and channel estimation is repeated for every subframe. The average BER results over all symbols and subframes are computed per antenna and the BER results per antenna are averaged again to obtain the final BER performance curves.  

The autoencoder transmitter incorporates perfect CSI information as well, directly into the encoding process as described in Section \ref{approach}.  As a result, the channel effect on the transmitted symbols is learned effectively during training, and a representation is learned to best preserve the information for the receiver.   

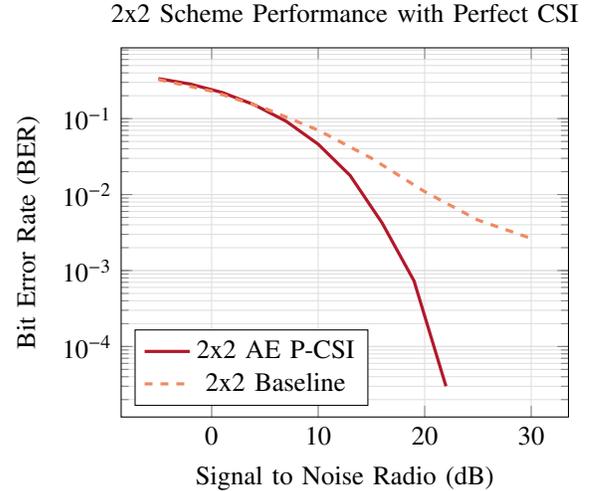
\begin{figure}[h]
\centering
\begin{tikzpicture}
  \begin{axis}[
            cycle list/RdGy-6,
            mark list fill={.!75!white},
            cycle multiindex* list={
                RdGy-6 \nextlist
                my marks \nextlist
                [3 of]linestyles \nextlist
                very thick \nextlist
            },  
	xmode=linear,
	ymode=log,
	xlabel=Signal to Noise Radio (dB), 
	ylabel=Bit Error Rate (BER),
	title={2x2 Scheme Performance with Perfect CSI},
	grid=both,
	minor grid style={gray!25},
	major grid style={gray!25},
	width=0.85\linewidth,
    legend pos=south west,
	no marks]
    \addplot table[x=snr,y=ber,col sep=comma]{2x2p.csv};
    \addplot table[x=snr,y=ber,col sep=comma]{2x2baseline.csv};
    \legend{{2x2 AE P-CSI},{2x2 Baseline}}
\addlegendentry{Baseline};
\end{axis}
\end{tikzpicture}
\caption{Error Rate Performance of Learned 2x2 Scheme (Perfect CSI).}
\label{fig:2x2berperf}
\end{figure}

In Figure \ref{fig:2x2berperf} we provide the performance curve measuring the BER from learned encoding scheme compared against the baseline for this perfect CSI case.  For both cases the total average BER results for all antennas are provided.  Here we obtain extremely promising results from the autoencoder approach as compared to the baseline method, potentially from some kind of inherent error correction scheme built in to the mapping and de-mapping process, as we have seen in \cite{intromlcomsys}.

\begin{figure}[h]
\centering
\includegraphics[width=0.5\textwidth]{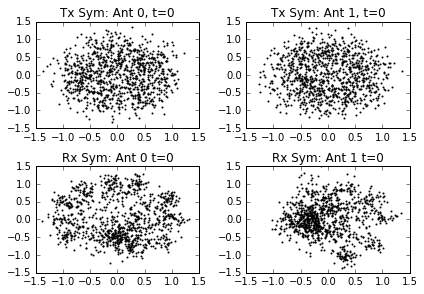}
\caption{Learned 2x2 Scheme, Random Channels.}
\label{fig:2x2consta}
\end{figure}

\begin{figure}[h]
\centering
\includegraphics[width=0.5\textwidth]{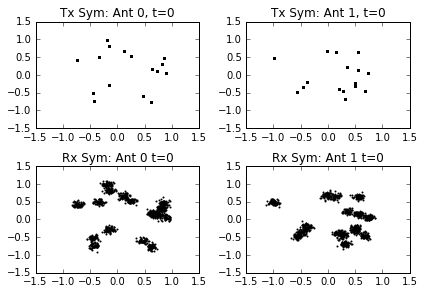}
\caption{Learned 2x2 Scheme, Diagonal Channel.}
\label{fig:2x2constb}
\end{figure}

\begin{figure}[h]
\centering
\includegraphics[width=0.5\textwidth]{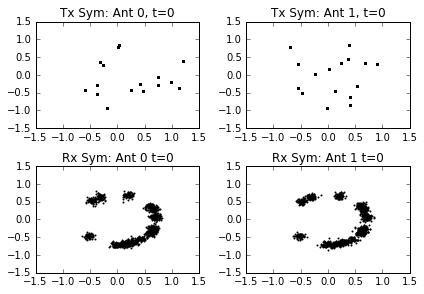}
\caption{Learned 2x2 Scheme, All-Ones Channel.}
\label{fig:2x2constc}
\end{figure}

Inspecting the transmit and receive constellations for the 2x2 scheme reveals an interesting structure that is formed.  In Figure \ref{fig:2x2consta} we show the constellations over a number of random channel draws, while in Figures \ref{fig:2x2constb} and \ref{fig:2x2constc} we show the receive constellations for the diagonal $\textbf{H}$ matrix and for an all-ones $\textbf{H}$ matrix.  Interestingly, for an $\textbf{H}$ matrix with roughly uniform power for each channel, we seem to obtain a receive waveform which is near constant amplitude phase encoding, while the transmit constellations appear to be quite random arrangements of $2^k = 16$ bits, forming a non-standard 16-QAM type arrangement.

\subsubsection{Quantized Channel Information at the Transmitter} \label{qcsi}

Our next step is to introduce feedback error in the simulation to reflect the practical implementation challenges. Since feedback of CSI, (e.g., from a handset device to a base station) for closed-loop schemes, requires protocol overhead in real world systems, we will consider the case where CSI is constrained to be compact discrete valued information encoded into a compact $v$-bit field.

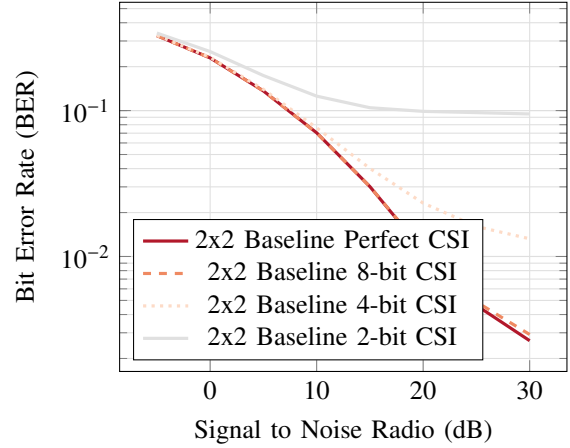
\begin{figure}[h]
\centering
\begin{tikzpicture}
  \begin{axis}[
            cycle list/RdGy-6,
            mark list fill={.!75!white},
            cycle multiindex* list={
                RdGy-6 \nextlist
                my marks \nextlist
                [3 of]linestyles \nextlist
                very thick \nextlist
            },  
	xmode=linear,
	ymode=log,
	xlabel=Signal to Noise Radio (dB), 
	ylabel=Bit Error Rate (BER),
	title={Baseline 2x2 Scheme Performance with Quantized CSI},
	grid=both,
	minor grid style={gray!25},
	major grid style={gray!25},
	width=0.85\linewidth,
    legend pos=south west,
	no marks]
    \addplot table[x=snr,y=ber,col sep=comma]{avgBER2bitsSNR-5to30dB1000sym100subframes2AntPaper.csv};
    \addplot table[x=snr,y=ber,col sep=comma]{avgBER8BitQuant2bitsSNR-5to30dB2AntPaper.csv};
    \addplot table[x=snr,y=ber,col sep=comma]{avgBER4BitQuant2bitsSNR-5to30dB2AntPaper.csv};
    \addplot table[x=snr,y=ber,col sep=comma]{avgBER2BitQuant2bitsSNR-5to30dB2AntPaper.csv};
    \legend{
        {2x2 Baseline Perfect CSI},
        {2x2 Baseline 8-bit CSI},
        {2x2 Baseline 4-bit CSI},
        {2x2 Baseline 2-bit CSI},
    }
\addlegendentry{Baseline};
\end{axis}
\end{tikzpicture}
\caption{Error Rate Performance of Baseline 2x2 Scheme (Quantized CSI).}
\label{fig:2x2baselineQuantized}
\end{figure}

We used LLoyd algorithm for quantizer implementation for the baseline MIMO system. Normally distributed random numbers were used to train the quantizer. The quantization values are known both at the transmitter and receiver in advance so that the receiver sends $v$-bit index of the quantized channel entries to the transmitter. System performance is very close to the perfect CSI case with 8-bit quantization but it goes down with decreasing number of quantization bits as shown in Figure \ref{fig:2x2baselineQuantized}.   

\begin{figure}[h]
\centering
\begin{tikzpicture}
  \begin{axis}[
            cycle list/RdGy-6,
            mark list fill={.!75!white},
            cycle multiindex* list={
                RdGy-6 \nextlist
                my marks \nextlist
                [3 of]linestyles \nextlist
                very thick \nextlist
            },  
	xmode=linear,
	ymode=log,
	xlabel=Signal to Noise Radio (dB), 
	ylabel=Bit Error Rate (BER),
	title={2x2 Scheme Performance With Quantized CSI},
	grid=both,
	minor grid style={gray!25},
	major grid style={gray!25},
	width=0.85\linewidth,
    legend pos=south west,
	no marks]
    \addplot table[x=snr,y=ber,col sep=comma]{2x2ae_quant_1bit.csv};
    \addplot table[x=snr,y=ber,col sep=comma]{2x2ae_quant_2bit.csv};
    \addplot table[x=snr,y=ber,col sep=comma]{2x2ae_quant_4bit.csv};
    \addplot table[x=snr,y=ber,col sep=comma]{2x2ae_quant_8bit.csv};
    \addplot table[x=snr,y=ber,col sep=comma]{2x2p.csv};
    \legend{
        {2x2 AE 1 Bit},
        {2x2 AE 2 Bit},
        {2x2 AE 4 Bit},
        {2x2 AE 8 Bit},
        {2x2 AE P-CSI}
    }
\addlegendentry{Baseline};
\end{axis}
\end{tikzpicture}
\caption{Error Rate Performance of Learned 2x2 Scheme (Quantized CSI).}
\label{fig:2x2Quantized}
\end{figure}
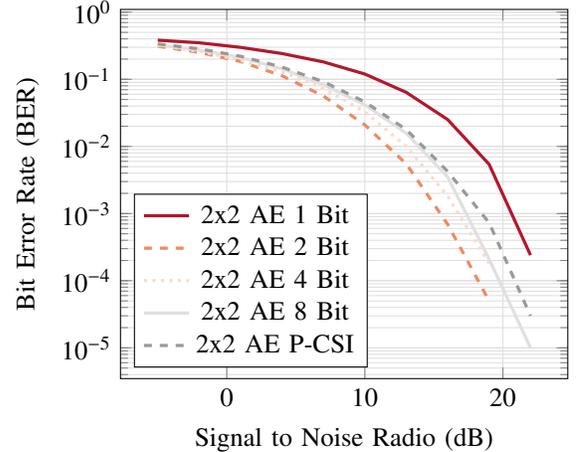

Inspecting while the BER curve results for the baseline method in Figure \ref{fig:2x2baselineQuantized} decrease in performance as we represent CSI with fewer bits, for the autoencoder case in Figure \ref{fig:2x2Quantized}, we actually see improved performance for some of the quantized CSI cases (2, 4 and 8 bit CSI) over the perfect CSI case. We believe that quantizing the CSI into $2^v$ discrete modes actually helps the learned system converge more easily to a good and well defined encoder configuration for each channel mode.  This is likely a much simpler target manifold to fit rather than attempting to fit the full complex mapping from real valued $\textbf{H}$ to some corresponding set of real valued encoder modes and it seems to provide a better solution which provides improved BER across a wide range of SNR levels while at the same time lending itself well to over the air deployment and use in future wireless standards with compact protocol overhead.

\begin{figure}[h]
\centering
\includegraphics[width=0.5\textwidth]{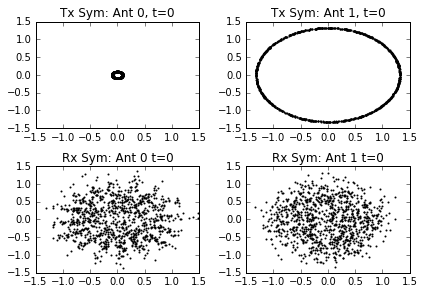}
\caption{Learned 2x2 Scheme 1 bit CSI Random Channels.}
\label{fig:2x2const_1ba}
\end{figure}

\begin{figure}[h]
\centering
\includegraphics[width=0.5\textwidth]{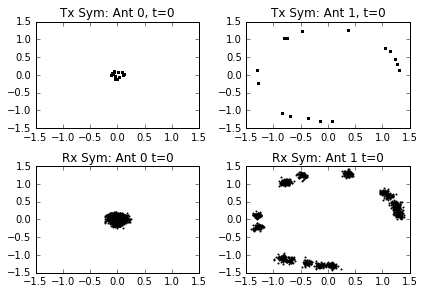}
\caption{Learned 2x2 Scheme 1-bit CSI All-Ones Channel.}
\label{fig:2x2const_1bo}
\end{figure}

The constellations learned under this scheme are interesting, as well.  For the 1-bit CSI case shown in Figure \ref{fig:2x2const_1ba} and \ref{fig:2x2const_1bo} we see that it learns a multi-level scheme, where antenna 0 and 1 transmit constant modulus encodings at two distinct power levels.  

\begin{figure}[h]
\centering
\includegraphics[width=0.5\textwidth]{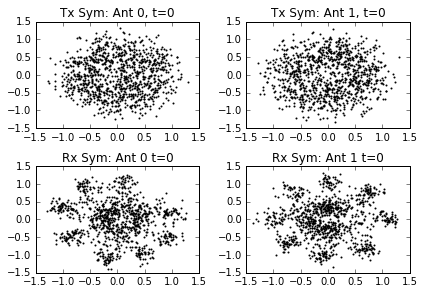}
\caption{Learned 2x2 Scheme 2 bit CSI Random Channels.}
\label{fig:2x2const_2ba}
\end{figure}

\begin{figure}[h]
\centering
\includegraphics[width=0.5\textwidth]{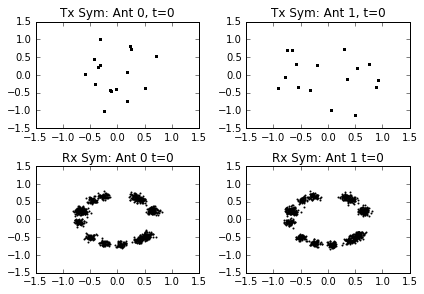}
\caption{Learned 2x2 Scheme 2 bit CSI All-Ones Channel.}
\label{fig:2x2const_2bo}
\end{figure}

For the 2-bit CSI case, shown in Figures \ref{fig:2x2const_2ba} and \ref{fig:2x2const_2bo} we can see that it learns a complex multi-level transmission scheme, some kind of irregular 16-QAM on each transmitter, where the receive copy, for roughly equal power paths turns out to be constant modulus.  

\subsubsection{Best Approach}

\begin{figure}[h]
\centering
\begin{tikzpicture}
  \begin{axis}[
            cycle list/RdGy-6,
            mark list fill={.!75!white},
            cycle multiindex* list={
                RdGy-6 \nextlist
                my marks \nextlist
                [3 of]linestyles \nextlist
                very thick \nextlist
            },  
	xmode=linear,
	ymode=log,
	xlabel=Signal to Noise Radio (dB), 
	ylabel=Bit Error Rate (BER),
	title={Overall Best 2x2 Scheme Performance},
	grid=both,
	minor grid style={gray!25},
	major grid style={gray!25},
	width=0.85\linewidth,
    legend pos=south west,
	no marks]
    \addplot table[x=snr,y=ber,col sep=comma]{2x2ae_quant_2bit.csv};
    \addplot table[x=snr,y=ber,col sep=comma]{2x2baseline.csv};
    \legend{
        {2x2 AE 2 Bit},
        {2x2 Baseline}
    }
\addlegendentry{Baseline};
\end{axis}
\end{tikzpicture}
\caption{Error Rate Performance of Best 2x2 Scheme.}
\label{fig:2x2best}
\end{figure}
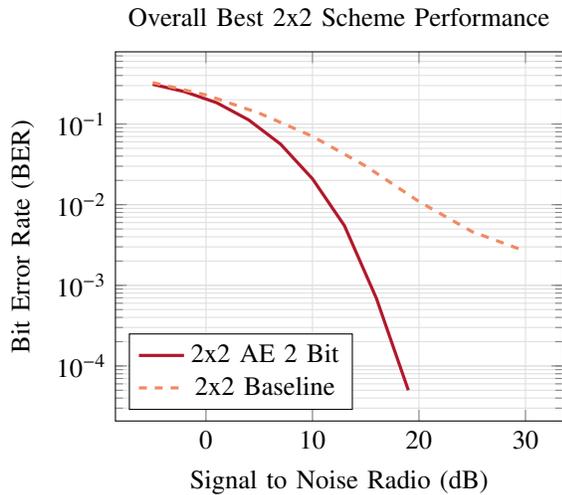

To summarize our findings for the 2x2 transmission scheme, we illustrate the best autoencoder approach among those evaluated in Figure \ref{fig:2x2best} and compare to the baseline performance.  We observe significant improvement of performance across the entire range of operation, with BER performance reaching below $0.5e{-5}$ below and SNR of 20dB.   We believe this is a powerful result which holds enormous promise for use in future wireless cellular and local area network systems, providing an extremely compact way to obtain a robust MIMO encoding scheme which jointly provides a strong degree of error correction and a significantly improved bit error rate performance compared to current day baselines.

\section{Discussions} \label{discussion}

The idea of learning a universal MIMO scheme for effective MIMO encoding across a range of different assumptions on CSI information, time-slots, antenna counts, and information densities is really quite appealing.  In our results, we have shown that the proposed method is quite competitive for 2x1 and 2x2 size channels without CSI, with perfect CSI, and especially with compact binary CSI.  We have also shown how such a system can be readily partitioned into a real world distributed communications system in order to efficiently manage CSI requirements.

We have seen that discretized CSI with this scheme works incredibly well, learning very compact CSI encodings and actually seems to help the autoencoder converge more rapidly and to a better general solution (given sufficient bits).  We have observed that learned solutions often favor constant modulus power at the receiver, but that in some cases they have learned sub-optimal solutions which split power unevenly between the two antennas even in the case of no-CSI.  

This work opens up numerous new avenues for investigation, many of which we hope to pursue in the near future. Among these, are combining $\hat{H} \mapsto H_v$ estimation with $y \mapsto H_v $  estimation, or in general allowing the channel estimation routines required to support this system to be learned, and with some error rather than directly using error free values of $H$. Similar analysis have already been performed for conventional single user MIMO systems as in \cite{ulukus}.  

Also, as many standards such as LTE \cite{LTEMIMO} and WiMaX define codebook indices to be transmitted compactly to convey CSI, we wish to provide additional comparison with these baselines, to determine of learned ``codebooks'' over- or under-perform the methods which are widely used today for closed-loop MIMO systems.

Additionally, the work needs to continue to evaluate performance on larger-scale MIMO arrangements such as 8x4 arrangements and massive MIMO, and to consider the case of multi-tap delay spreads.  The work has been conducted in the time-domain rather than in the conventional long-symbol MIMO/OFDM domain, and so numerous connections may be drawn still in connecting these two domains optimally.

Finally, we will extend the MIMO described here in the single user case to multi-user (MU-MIMO) which includes Multiple Access; i.e., $N$ transmitters to $1$ receiver, and Broadcast; i.e. $1$ transmitter to $N$ receivers, channels. Current schemes for these MIMO problems are known to have sub-optimal implementations due to complexity issues in current day systems such as the implementation of capacity-achieving Dirty Paper Coding for Broadcast channel \cite{costa}.  We believe that a learning based approach holds enormous promise in tackling and gaining from this complexity in an elegant and manageable way.

\section{Conclusion} \label{conclusion}

The architecture and initial results described herein provide an exciting new approach to physical layer design and optimization for MIMO wireless communications schemes using an autoencoder.  This is a significant departure from current day systems, and currently has many issues which still need to be understood and made to operate efficiently with real world physical constraints.  However, preliminary results show that performance can be highly competitive with existing schemes and holds the promise of doing so at much lower computation complexity using wide-concurrent implementations.  We hope that our results help to illustrate the power of the autoencoder approach to communications in learning to optimize representations and encoding/decoding processes for complex channel regimes and data.  We hope that as results and findings continue to increase in maturity in this field, this naive, approximate approach to signal processing will continue to gain increasing acceptance and understanding throughout the wireless communications and signal processing field.  This approach makes it difficult to provide rigorous analytic performance bounds and guarantees under the current widespread understanding of deep learning and ability to apply analytic methods at scale.  While these analytic properties are desirable, and forgoing them has not proved popular with all in the signal processing community, it is sometimes helpful to recall the words of George Box who wrote elegantly, "All models are wrong but some are useful" \cite{box1976science}.  

\printbibliography

\end{document}